\documentclass[a4paper]{article}

\usepackage{INTERSPEECH2022}
\usepackage{multirow}
\usepackage{algorithm}
\usepackage{algpseudocode}
\usepackage{dblfloatfix}

\title{Design Guidelines for Inclusive Speaker Verification Evaluation Datasets}
\name{Wiebke (Toussaint) Hutiri$^1$, Lauriane Gorce$^2$, Aaron Yi Ding$^1$}
\address{
  $^1$Delft University of Technology, 
  $^2$Open North}
\email{w.toussaint@tudelft.nl, lauriane@opennorth.ca, aaron.ding@tudelft.nl}

\begin{document}

\maketitle
\begin{abstract}

Speaker verification (SV) provides billions of voice-enabled devices with access control, and ensures the security of voice-driven technologies. As a type of biometrics, it is necessary that SV is unbiased, with consistent and reliable performance across speakers irrespective of their demographic, social and economic attributes. Current SV evaluation practices are insufficient for evaluating bias: they are over-simplified and aggregate users, not representative of usage scenarios encountered in deployment, and consequences of errors are not accounted for. This paper proposes design guidelines for constructing SV evaluation datasets that address these short-comings. We propose a schema for grading the difficulty of utterance pairs, and present an algorithm for generating inclusive SV datasets. We empirically validate our proposed method in a set of experiments on the VoxCeleb1 dataset. Our results confirm that the count of utterance pairs/speaker, and the difficulty grading of utterance pairs have a significant effect on evaluation performance and variability. Our work contributes to the development of SV evaluation practices that are inclusive and fair.

  
\end{abstract}
\noindent\textbf{Index Terms}: speaker verification, voice biometrics, evaluation, audit, bias, fairness, design guidelines

\section{Introduction}
\label{s:introduction}

Speaker verification (SV) technology, which determines the identity of a speaker from their voice, is widely used across the world as a form of biometric identification~\cite{Kinnunen2009Overview}. Applications range from devices like mobile phones and smart speakers to transactional, investment and mobile banking, call centers and proof-of-life verification of pensioners. While unobtrusive, SV is extremely sensitive to the recording hardware, the deployment context and environment~\cite{Reynolds2002Overview}. Moreover, it is also sensitive to the demographic attributes of speakers~\cite{hansen2015speaker}.

In machine learning (ML), bias has become a priority concern for fair and inclusive development of ML technologies. A ML system is considered biased if its predictive outputs favour or are prejudiced against individuals or groups of people based on their demographic attributes~\cite{mehrabi2019survey}. An often-times unintended consequence of bias is discrimination, which can lead to public mistrust~\cite{toreini2020relationship} and legal consequences~\cite{wachter2021bias}. It is thus important that potenatial bias in SV systems is evaluated and mitigated during development. Studies of bias in SV, and research to mitigate its effects, are currently limited. Recent research has highlighted that existing SV evaluation practices contribute to exacerbating bias in the development of SV systems~\cite{Toussaint2022Bias}. In particular, unrepresentative benchmarks make it impossible to evaluate bias, let alone address it.

This paper presents design guidelines for the creation of more inclusive evaluation datasets that are suitable for evaluating bias in SV. While existing SV datasets consider demographic representation on a speaker level~\cite{Nagrani2020a, Qin2020himia, mclaren2016speakers}, we show that it is also important to consider inclusion at the more fine-grained level of utterance pairs. We address two key questions: 
\begin{enumerate}
    \item How should utterance pairs be constructed to support an unbiased evaluation?
    \item How many utterance pairs are needed per speaker for a robust SV evaluation?
\end{enumerate} 
Using an iterative design approach, we propose an algorithm for constructing evaluation datasets that are representative of realistic application scenarios. We empirically validate the utility of the datasets. Our experimental results present evidence that the difficulty of utterance pairs impacts the evaluation outcome. We also show that randomized utterance pairings can result in significant performance variation if the utterance pair count per speaker is low. Based on these results we propose design guidelines that consider inclusion on an utterance level.

In the next section we present background information on SV evaluation. Design considerations for the evaluation dataset are discussed in Section~\ref{s:dataset_design_considerations}. Section~\ref{s:experiments} outlines the experimental setup and results. We propose evaluation dataset design guidelines and an outlook on future work in Section~\ref{s:guidelines} and conclude in Section~\ref{s:conclusion}.

\section{Background}
We now present a short overview of SV evaluation, and discuss shortcomings of current practices. For a detailed overview of SV techniques and systems we refer the reader to~\cite{Kinnunen2009Overview} for a historic, and to~\cite{Bai2021Speaker} for a recent survey of deep learning methods.

\subsection{Speaker Verification Evaluation}

SV performance is determined by the false positive (FP) rate and false negative (FN) rate at a threshold value to which the system has been calibrated. It is accepted that the two error rates present a trade-off, and that selecting an appropriate threshold is an application-specific design decision~\cite{NIST2020}. The threshold value is determined by balancing the FP and FN error rates for a particular objective, such as obtaining an \emph{equal error rate} (EER) for FP and FN errors, or minimising a cost function. The US National Institute of Standards and Technology (NIST), which has been conducting speaker recognition evaluations for over two decades, advises against the use of the \emph{EER}, as applications typically require either low FP or FN~\cite{greenberg2020two}. The \emph{detection cost function} (DCF) is a weighted sum of FP and FN rates across threshold values, with weights determined by the application requirements~\cite{greenberg2020two}. To compare performance across models, systems are frequently tuned to the threshold value at the minimum of the DCF (\emph{minDCF}), and the \emph{minDCF} is reported as metric. NIST recommends that SV systems are evaluated with Detection Error Tradeoff (DET) curves to show SV performance across a range of error rates. 

\subsection{Short-comings of Current Evaluation Practices}
Alongside the NIST, independent SV competitions are a popular mechanism for evaluating and driving advancement in SV techniques. Many competitions, such as SITW~\cite{mclaren2016speakers}, the VoxCeleb SRC~\cite{Nagrani2020Voxsrc} and the Far-Field SRC~\cite{qin2020interspeech} have been hosted at Interspeech. The competitions serve as benchmarks, promote evaluation habits and implicitly guide the culture of evaluation in the domain. To enable comparison across competition entries, they focus their evaluations on one or two metrics, typically a version of the \emph{minDCF} or \emph{EER}.

Recent research on bias in SV has shown that evaluation outcomes are highly dependent on the evaluation dataset and metrics used for evaluation~\cite{Toussaint2022Bias}. The paper found that current SV evaluation practices consider limited context, environment and user variability in the design of evaluation datasets and metrics. The authors argue that SV evaluation practices present an oversimplified view of application scenarios, promote unrepresentative benchmark populations and oversimplified metrics, do not consider the consequences of errors and favour low FP rates while neglecting the consequences of FN rates. This limits current evaluation practices in their ability to detect bias in realistic application settings and adds to bias that may already exist in SV technologies. We build on this work by studying approaches to mitigating bias in the design of evaluation datasets to ensure that evaluation practices are more inclusive and fair. 




\section{Dataset Design Considerations}
\label{s:dataset_design_considerations}

In this section we consider consequences of speaker verification (SV) errors, introduce a schema for grading utterance pairs, and propose an algorithm for generating inclusive SV evaluation datasets. 

We use the following terminology and notation. A \emph{speaker} ($S^i$) is a unique individual with at least one speech \emph{utterance} ($u^i$) in the set of all \emph{evaluation utterances} ($\mathcal{U}$). $\mathcal{U}^i$ is the subset of evaluation utterances of speaker $S^i$ and $S=\{S^1...S^K\}$ is the set of all speakers in $\mathcal{U}$. An \emph{evaluation dataset} ($\mathcal{D}$) is constructed by generating utterance pairs $\{u_a, u_b\}$ from the evaluation utterances in $\mathcal{U}$. Conventionally $u_a$ is the enrollment utterance, and $u_b$ is the test utterance. Practically, the order of utterance pairs does not matter in SV evaluation. 

A \emph{same speaker} pair $\{u^1_a, u^1_b\}$ has the enrollment and test utterances drawn from $\mathcal{U}^1$, while a \emph{different speaker} pair $\{u^1_a, u^2_a\}$ has the enrollment and test utterances drawn from two different speakers, ie from $\mathcal{U}^1$ and $\mathcal{U}^2$. The objective of creating an inclusive SV evaluation dataset, $\mathcal{D}$, is then to:

\begin{itemize}
\itemsep0em 
    \item offer an \textbf{equivalent evaluation} across all speakers in $S$
    \item generate same speaker and different speaker utterance pairs that are \textbf{reflective of deployment usage scenarios}
    \item facilitate an evaluation that is \textbf{robust to perturbations} in utterance pairings (e.g., if $\{u^1_a, u^2_a\}$ is substituted with $\{u^1_b, u^3_f\}$ the false positive error rate should not change significantly)
\end{itemize}

\subsection{Difficulty Grading of Utterance Pairs}

SV is a comparative task that compares an enrolled speaker utterance with a test utterance based on the difference between embeddings of the voice profiles. Naturally, system performance is affected by the utterance pairs being compared. However, not all utterance pairs are born equal. 

It is important that evaluation utterance pairs test likely scenarios that a SV system will encounter when deployed, and that pairs of appropriate difficulty are constructed to test the limits of the system. A qualitative difficulty grading of utterance pairs has the advantage that it bares no computational cost and that it can be interpreted in relation to the application context. We thus created a schema for grading the difficulty of utterance pairs across four categories: trivial (cat 1), easy (cat 2), medium (cat 3) and hard (cat 4) for same and different speaker pairs. As a general rule, evaluation of same speakers is easier if utterance pairs are more similar, and harder if utterance pairs are more different. For different speakers the opposite is true: similar utterance pairs are harder, and different utterance pairs are easier to classify. Speaker, recording channel and environmental attributes affect the similarity of utterance pairs. In Table~\ref{tab:utterance_pair} we show how we use the schema to grade utterance pair difficulty based on speaker gender, nationality, recording channel and background noise (attributes chosen based on available metadata in the VoxCeleb1 dataset). Same speaker pairs always have the same gender and nationality, while different speaker pairs are never from the same recording. We assume that utterances from the same recording have the same noise conditions. 

\begin{table}[hbt]
\centering
\footnotesize
\begin{tabular}{p{0.11\linewidth}p{0.18\linewidth}|p{0.07\linewidth}p{0.12\linewidth}p{0.11\linewidth}p{0.07\linewidth}}
Utterance\newline Pairs& Difficulty & Same\newline Gender & Same\newline Nationality & Same\newline Recording & Same\newline Noise\\ \midrule 
Same & cat 1 (trivial) & Yes & Yes & Yes & Yes \\
Speaker & cat 3 (med.) & Yes & Yes & No & n.k.\\ \midrule 
 & cat 1 (trivial) & No & No & No & n.k.\\
Different & cat 2 (easy) & No & Yes & No & n.k.\\
Speakers & cat 3 (med.) & Yes & No & No & n.k.\\
 & cat 4 (hard) & Yes & Yes & No & n.k.
\end{tabular} \smallskip
\caption{Grading of utterance pairs (n.k. = not known)}
\label{tab:utterance_pair}
\vspace{-5mm}
\end{table}

\noindent

\subsection{Generating Evaluation Datasets}

Taking these considerations and the dataset requirements into account, we designed Algorithm~\ref{alg:eval} for generating an evaluation dataset, $\mathcal{D}$, from a set of utterances, $\mathcal{U}$. Given available utterances, our intention was to create an evaluation dataset in the most difficult category. To ensure that the evaluation is robust to perturbations in utterance pairings, the random seed for selecting same and different speaker utterance pairs can be changed to generate similar datasets with different utterance pairings. SV evaluation outcomes can then be compared across the datasets.


\begin{algorithm}
\footnotesize
\caption{Evaluation Dataset Generation Algorithm}\label{alg:eval}
\begin{algorithmic}
\State $S \gets$ set of all speakers
\State $i \gets$ unique speaker
\State $a, b \gets$ two recording instances
\State $u \gets$ utterance
\State $\mathcal{U} \gets$ set of all utterances
\State $n \gets$ count of \emph{different speaker} utterance pairs/speaker
\State $r \gets$ random seed
\State $\mathcal{D} \gets$ new evaluation dataset
\For{$S^i$ in $S$}
    \State $\mathcal{U}^i \gets $ subset of $u^i$
    \State $\mathcal{U}^j \gets $ subset of $u^j$ \Comment{$i\neq j$; $S^i, S^j$ same group}
    \State $\mathcal{D}^i_{same} \gets $ new list of \emph{same speaker} utterance pairs 
    \State $\mathcal{D}^i_{diff} \gets $ new list of \emph{different speaker} utterance pairs
    \For{$u^i_a$, $u^i_b$ in $\mathcal{U}^i$} \Comment{\textbf{same speaker pairs}}
        \If{recording$_a$ is not recording$_b$} \Comment{difficulty: medium}
        \State $\mathcal{D}^i_{same}$ append ${\{u^i_a, u^i_b\}}$
        \Else{}
        \State pass
        \EndIf
    \EndFor
    \State $\mathcal{D}^i_{same} \gets $ randomly select $n$ pairs from $\mathcal{D}^i_{same}$ with seed $r$
    \State $\mathcal{U}^i_n \gets $ randomly select $n$ $u$ from $\mathcal{U}^i$ with seed $r$ (replace ok)
    \State $\mathcal{U}^j_n \gets $ randomly select $n$ $u$ from $\mathcal{U}^j$ with seed $r$
    \State $x \gets 0$
    \While{$x < n$}
        \For{$u^i_x$, $u^j_x$ in $\mathcal{U}^i_n$, $\mathcal{U}^j_n$} \Comment{\textbf{different speaker pairs}}
            \State $\mathcal{D}^i_{diff}$ append ${\{u^i_x, u^j_x\}}$
        \EndFor
    \EndWhile
    \State $\mathcal{D}$ append $\mathcal{D}^i_{same}$
    \State $\mathcal{D}$ append $\mathcal{D}^i_{diff}$
\EndFor
\end{algorithmic}
\end{algorithm}
\section{Experiments}
\label{s:experiments}

We empirically validated our approach to generating SV evaluation datasets by conducting experiments with the VoxCeleb1 dataset~\cite{Nagrani2017Voxceleb}. Due to space limitations we only show experiments and results for speaker groups based on nationality. 
Our experiments show that model performance deteriorates significantly when evaluated on datasets with more difficult utterance pairs of same and different speakers. Moreover, when the number of utterance pairs per speaker is low, model performance has high variability. Evaluation robustness improves as the number of utterance pairs increases. In this section we describe the setup of the experiments and the baseline evaluation dataset, and then present results to demonstrate the effect of utterance pair grading and count.

\subsection{Setup}
We use a pre-trained end-to-end model based on a 34-layer ResNet trunk architecture. The model is described in detail in~\cite{heo2020clova} where it is referred to as the performance optimized model. The model has been trained on the VoxCeleb~2 training set~\cite{Nagrani2020a}, with close to 1 million utterances of 5994 speakers. 61\% of speakers are male and 29\% of speakers have a US nationality, which is the most represented nationality. The feature input to the model are 64-dimensional log Mel-filterbanks.

\noindent
We use the model in a SV inference pipeline to classify utterance pairs in an evaluation dataset $\mathcal{D}$ as being from the same or from different speakers. The only variable that we consider in the experiments is $\mathcal{D}$. Evaluation datasets were generated from VoxCeleb1 with the algorithm described in Section~\ref{s:dataset_design_considerations}. Different speaker pairs were constructed from speakers with the same gender and nationality. The evaluation sets are characterised by speaker nationalities, the count of utterance pairs/speaker~($n$)\footnote{$n$ is the count of same \emph{or} different utterance pairs. As these two counts are equal, the total number of utterance pairs per speaker is $2n$.} and the random seed with which the dataset was generated. The experiments are detailed in Table~\ref{tab:experiments}.

\begin{table}[hbt]
\footnotesize
\centering
\begin{tabular}{p{0.04\linewidth}p{0.35\linewidth}p{0.25\linewidth}p{0.16\linewidth}}
\textbf{Exp} & \textbf{Nationalities} \newline (speaker count) & \textbf{Pairs/speaker\newline ($n$)} & \textbf{Random\newline seed} \\ \midrule
1 & Canada, India, USA, Ireland, Norway, UK, Australia, Germany & 520 & 12 \\ \midrule
2 & Canada (54), India (26), UK (215)  & 50, 100, 150, 225, 350, 450, 520 & 3, 6, 8, 12, 20 \\
\end{tabular} \smallskip
\caption{Overview of SV evaluation experiments.}
\label{tab:experiments}
\vspace{-8mm}
\end{table}

\subsection{Baseline Evaluation Dataset}

To benchmark our evaluation datasets, we compare them against a baseline evaluation on VoxCeleb1-H. Of the three VoxCeleb1 evaluation sets, VoxCeleb1-H is considered the "hard" set. It includes only different speaker pairs of the same gender and nationality, and requires groups of 5 or more speakers. Table~\ref{tab:voxceleb_same_speaker_utterance_pairs} summarises the attributes of same speaker pairs in VoxCeleb1~H. As can be seen from the table, the number of speakers, the total number of utterance pairs, and the count of utterances pairs per speaker varies across speaker groups based on nationality. Additionally, all groups contain trivial (cat 1) same speaker pairs. However, the proportion of trivial pairs in relation to all of the group's same speaker pairs differs. 

\begin{table}[hbt]
\footnotesize
\centering
\begin{tabular}{lclp{0.15\linewidth}p{0.15\linewidth}}
\textbf{Nationality} & \textbf{Speakers} & \textbf{Pairs} & \textbf{Pairs/\newline speaker} & \textbf{cat 1\newline (trivial)} \\ \midrule
USA & 799 & 178122 & 222.9 & 12.9\% \\
UK & 215 & 53111 & 247.0 & 10.3\% \\
Canada & 54 & 10864 & 201.2 & 11.1\% \\
India & 26 & 10053 & 386.7 & 10.6\% \\
Australia & 37 & 8668 & 234.3 & 10.5\% \\
Ireland & 18 & 4960 & 275.6 & 8.5\% \\
Norway & 20 & 4906 & 245.3 & 10.0\% \\
New Zealand & 6 & 1811 & 301.8 & 10.1\% \\
Germany & 5 & 1256 & 251.2 & 17.0\% \\
Mexico & 5 & 1130 & 226.0 & 10.2\% \\
Italy & 5 & 571 & 114.2 & 17.0\%
\end{tabular} \smallskip
\caption{VoxCeleb1-H Same Speaker Utterance Pairs.}
\label{tab:voxceleb_same_speaker_utterance_pairs}
\vspace{-5mm}
\end{table}


\subsection{Results}
\label{s:results}

We now analyze the results of our experiments and show how utterance pair grading and utterance pair count affect the evaluation outcome. 

\begin{figure}[hbt]
    \centering
    \includegraphics[width=\linewidth]{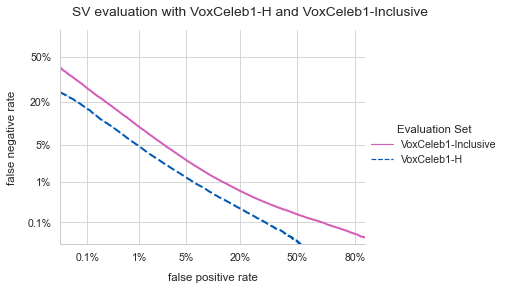}
    \caption{DET curves for a baseline and our inclusive dataset show significant performance decline when the latter is used for evaluation. This implies that the SV model performs worse under more realistic and inclusive evaluation scenarios.}
    \label{fig:det_curves_new_dataset}
    \vspace{-5mm}
\end{figure}

\begin{figure*}[!t]
\centering
    \includegraphics[width=\textwidth]{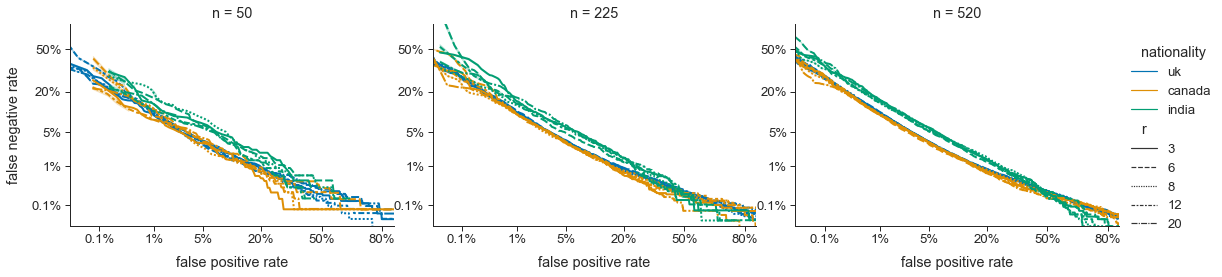}
    \caption{DET curves show variability in evaluation outcomes for evaluation sets with 50, 225 and 520 utterance pairs ($n$) for Canadian, Indian and UK speakers. For each $n$ five datasets were generated with different random seeds.}
    \label{fig:det_curves_utterance_pairs}
    \vspace{-3mm}
\end{figure*}

\subsubsection{Effect of Utterance Pair Grading}

Experiment 1 included speakers from 8 nationalities, with 520 utterance pairs of different speakers (1040 total pairs) per speaker. We call this dataset \emph{VoxCeleb1-Inclusive}. The key differences between the VoxCeleb1-H baseline and VoxCeleb1-Inclusive are that the latter contains an equal count of pairs/speaker across individual speakers and speaker groups, that trivial same speaker pairs have been excluded and that speakers were included if their unique utterance pair count for different speakers was greater than our selection value $n$. Due to our randomized utterance pair generation, the different speaker pairs in the two evaluation sets are not the same.

\noindent
The DET performance curves for the model evaluated on the two datasets is shown in Figure~\ref{fig:det_curves_new_dataset}. When evaluated on VoxCeleb1~H performance is reasonable: a FP rate of 1\% results in a FN rate of 5\%. However, at the same FP rate the FN rate triples to ~15\% when same speaker utterance pairs are exclusively of medium difficulty grading.

\subsubsection{Effect of Utterance Pair Count}

In Experiment 2 we investigate the effect of the utterance pair count on the robustness of the evaluation, given perturbations in utterance pairings. We compare the evaluation outcome for Canadian, Indian and UK speakers across evaluation datasets generated with 50, 100, 150, 225, 350, 450 and 520 distinct different speaker utterance pairs/speaker. For each utterance pair count $n$ we generated five versions of the dataset with different random seeds. 

\begin{figure}[!hbt]
    \centering
    \includegraphics[width=\linewidth]{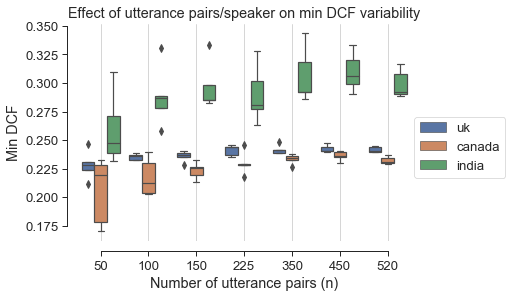}
    \caption{Variance in the minDCF metric for Canadian, Indian and UK speakers for evaluation datasets with different counts of utterance pairs/speaker ($n$). Five versions of each dataset are compared. The evaluation is more robust when $n$ is larger.}
    \label{fig:mindcf_utterance_pairs}
    \vspace{-3mm}
\end{figure}

\noindent
For the equal error rate (\emph{EER}) and minimum Detection Cost Function (\emph{minDCF}) performance metrics we observe significant variability in performance when $n$ is small. Results for the \emph{minDCF} are shown in Figure~\ref{fig:mindcf_utterance_pairs}. To illustrate, the \emph{minDCF} of the worst performing evaluation is 30\% higher than the best evaluation for Canadian speakers when $n=50$. This performance difference can be attributed entirely to utterance pairings generated with different random seeds. In general, we observe that variance is greater when the count of unique speakers in a group is smaller: UK speakers experience the lowest variance, while Indian speakers experience the greatest variance. 

Variability is not limited to performance on the \emph{EER} and \emph{minDCF} metrics, but exists across the DET performance curve, as shown in Figure~\ref{fig:det_curves_utterance_pairs}. As $n$ increases, variability remains greatest at the end points of the curve where FP and FN error rates are the greatest/smallest. Variability stablilizes first in the center of the curve. This means that at very low FP rates the expected FN rate will carry significant uncertainty (and vice versa), unless $n$ is large enough to ensure stable performance. 

\section{Evaluation Dataset Design Guidelines}
\label{s:guidelines}

Based on our analysis and empirical validation, we propose that an inclusive evaluation dataset for robust speaker verification evaluation should:
\begin{enumerate}
\itemsep0em 
    \item have an equal number of same speaker and different speaker utterance pairs for each speaker
    \item have at least 500 different speaker utterance pairs for each speaker
    \item have an equal number of utterance pairs per speaker 
    \item have an equal proportion of utterance pairs of each difficulty grading per speaker
    \item have utterance pairs with difficulty gradings that are representative of usage scenarios of deployed applications
    \item be compared against datasets variations with randomly generated utterance pairings to ensure robust evaluation outcomes
\end{enumerate}

\noindent
\textbf{Limitations and Future Work}: 
While we studied the effects of utterance pairs on SV evaluation outcomes, we did not conduct a detailed study on the number of unique speakers that are required for a robust evaluation. This is an opportunity to extend the proposed design guidelines in future research. Our classification of utterance pairs by difficulty grading was limited by the metadata available for the VoxCeleb1 dataset. In actual usage scenarios hard utterance pairs may be more challenging than the pairs we evaluated. For example, smart device applications should consider different utterance pairs from same gender family members or colleagues, which present the most likely intruders to these systems. Future work should also consider speaker groups along different dimensions, such as age, accent, health conditions and time.

\section{Conclusion}
\label{s:conclusion}

In this paper we present design guidelines for generating inclusive evaluation datasets towards robust speaker verification (SV) evaluation. During an iterative development process we elaborate on design considerations that account for consequences of SV errors, propose a taxonomy for grading the difficulty of utterance pairs, and present an algorithm for generating inclusive evaluation datasets. We empirically validate the evaluation datasets generated with our proposed method in a set of experiments on the VoxCeleb1 dataset. Our results confirm that the count of utterance pairs/speaker, and the difficulty grading of utterance pairs have a significant effect on evaluation outcomes. We advocate that current speaker verification evaluation practices can become more inclusive and robust by including equal counts of utterance pairs with the same difficulty grading for all individual speakers and groups.  





\bibliographystyle{IEEEtran}

\bibliography{references}

\end{document}